\documentclass{aa}
\usepackage{txfonts}
\usepackage{psfig}
\usepackage{wasysym}
\usepackage{color}

\begin{document}

\newcommand{\beqa}{\begin{eqnarray*}}
\newcommand{\eeqa}{\end{eqnarray*}}
\newcommand{\beqan}{\begin{eqnarray}}
\newcommand{\eeqan}{\end{eqnarray}}
\newcommand{\beq}{\begin{equation}}
\newcommand{\eeq}{\end{equation}}
\newcommand{\diff}{{\rm d}}
\newcommand{\drr}{\frac{\partial}{\partial r}}
\newcommand{\dtt}{\frac{\diff}{\diff t}}
\newcommand{\dr}[1]{\frac{\partial  #1}{\partial r}}
\newcommand{\dt}[1]{\frac{\partial  #1}{\partial t}}
\newcommand{\lp}{ \left(}
\newcommand{\rp}{ \right)}
\newcommand{\lc}{ \left[}
\newcommand{\rc}{ \right]}
\newcommand{\cf}{{\it cf.}}
\newcommand{\ie}{{\it i.e.}}
\newcommand{\eg}{{\it e.g.}}
\newcommand{\Suzanne}[1]{\texttt{\textbf{{[#1]}}}}
\newcommand{\CC}[1]{\texttt{\textsl{\textbf{\color{red}{[#1]}}}}}

\renewcommand{\topfraction}{1.} 
\renewcommand{\textfraction}{0.} 

\def\O{\Omega}

\bibliographystyle{plain}

\title{Angular momentum transport by internal gravity waves}
\subtitle{IV - Wave generation by surface convection zone, from the pre-main sequence 
to the early-AGB in intermediate mass stars}

\author{Suzanne Talon\inst{1} and Corinne Charbonnel\inst{2,3}}

\offprints{Suzanne Talon}

\institute{
D\'epartement de Physique, Universit\'e de Montr\'eal, Montr\'eal PQ H3C 3J7, Canada
\and Geneva Observatory, University of Geneva, ch. des Maillettes 51, 1290 Sauverny, Switzerland
\and Laboratoire d'Astrophysique de Toulouse et Tarbes, CNRS UMR 5572, OMP, Universit\'e Paul
Sabatier 3, 14 Av. E.Belin, 31400 Toulouse, France \\
(talon@astro.umontreal.ca, Corinne.Charbonnel@obs.unige.ch)}

\date{Received / Accepted for publication in A\&A }

\authorrunning{S. Talon \& C. Charbonnel}
\titlerunning{Internal gravity waves from the PMS to the early-AGB}

  \abstract
   {This is the fourth in a series of papers that deal with angular momentum transport
   by internal gravity waves in stellar interiors.}
   {Here, we want to examine the potential role of waves in other evolutionary
   phases than the main sequence.}
   {We study the evolution of a $3\,M_\odot$ Population\,I model from the pre-main sequence
   to the early-AGB phase and examine whether waves can lead to angular momentum
   redistribution and/or element diffusion at the external convection zone boundary.}
   {We find that, although waves produced by the surface convection zone 
   can be ignored safely for such a star during the main sequence, it is not
   the case for later evolutionary stages. In particular, angular momentum transport by 
   internal waves could be quite important at the end of the sub-giant branch and during
   the early-AGB phase. Wave-induced mixing of chemicals is expected during the
   early-AGB phase.}
   {}


   \maketitle
%

\section{Introduction}

In recent years, several authors studied the impact of internal gravity waves (IGWs) 
in a variety of main sequence stars. These waves were initially invoked as a source
of mixing in stellar interiors in low-mass stars with 
an extended surface convection
zone (Press~1981; Garc\'\i a L\'opez \& Spruit~1991;
Schatzman~1993; Montalb\'an~1994) and also as an efficient process in the synchronization
of massive binary stars (Goldreich \& Nicholson~1989). More recently, it was suggested that
IGWs may play a role in braking the solar core 
(Schatzman~1993; Zahn, Talon, \& Matias~1997;
Kumar \& Quataert~1997). This idea was confirmed first in static models
(Talon, Kumar, \& Zahn~2002) and recently in the complete evolution of solar-mass models,
evolved all the way from the pre-main sequence to 4.6\,Gy (Charbonnel \& Talon~2005). 

All these authors find that IGWs are easily excited, 
and a similar conclusion is reached by
studies of convection on top of a stably stratified layer in 2--D and 3--D hydrodynamic
numerical simulations (\eg Hurlburt et al.~1986,~1994; 
Andersen~1994; Nordlund et al.~1996; Kiraga et al.~2003; Dintrans et al.~2005; Rogers \& 
Glatzmeier~2005a,b). These waves have a strong impact on stellar evolution,
especially through their effect on the rotation profile. 
Through differential filtering, IGWs indeed play a major role in the redistribution 
of angular momentum in stars,  which determines the extent and magnitude of 
rotation-induced mixing (Talon \& Charbonnel~2005, hereafter TC05). 

In the spirit of applying a unique set of physical
principles to stellar evolution, one needs to assess the impact of such waves on stars of
all masses and at various evolutionary stages.
It is our purpose in this series of papers to consistently examine the full  
Hertzsprung-Russel diagram (HRD) for determining when such waves are efficiently emitted and 
how they can affect stars when their complete rotational history is being
considered.

In Talon \& Charbonnel~(2003, hereafter Paper~I; see also Talon \& Charbonnel 1998), 
we showed how the appearance of IGWs in
solar-metallicity main sequence stars with an effective temperature 
$T_{\rm eff} \apprle 6700~{\rm K}$ (i.e., when the surface convection zone becomes 
substantial) can explain the existence of the 
lithium dip\footnote{The so-called lithium dip, observed in all galactic
open clusters older than $\approx 200~{\rm Gyr}$, as well as in the field, 
refers to a strong lithium depletion in a narrow
region of $\approx 300~{\rm K}$ in effective temperature, 
centered around $T_{\rm eff} = 6700~{\rm K}$
(e.g. Wallerstein et al.~1965; Boesgaard \& Tripicco~1986; Balachandran~1995).} 
in stars undergoing rotational mixing.

In Talon \& Charbonnel~(2004, hereafter Paper~II), we examined the 
IGW generation in Population~II main sequence stars. 
We showed that, along the lithium plateau\footnote{The lithium plateau refers 
to the constant Li abundance of Population~II dwarf stars above 
$T_{\rm eff} \approx 5500~{\rm K}$ (e.g. Spite \& Spite~1982).},
the net angular momentum luminosity of IGWs is constant 
and high enough to enforce quasi solid-body rotation similar to that
of the Sun in these stars.
We proposed that this behavior could play a major role in explaining
the constancy of the lithium abundance in the oldest 
dwarf stars of our Galaxy (see also a discussion in Charbonnel \& Primas~2005). 

Here, we wish to look at other evolutionary stages, especially the pre-main sequence 
(PMS) and the more advanced stages for intermediate-mass stars.
We focus in particular on stars in which IGWs generation
by the surface convection zone is limited on the 
main sequence, \ie, Pop\,I star originating from the left side of the Li dip. 
To do so, we follow the evolution of a $3\,M_\odot$, $Z=0.02$ star 
from the PMS up to the early asymptotic giant branch (early-AGB)\footnote{The peculiar case
of thermally pulsing AGB (TP-AGB) stars will be presented elsewhere.}.
We estimate at which stages waves are efficiently generated in the outer convection zone. 
This is a prerequisite for evaluating 
their impact on the rotation profile of the corresponding stellar models.
We also look at the existence of a shear layer oscillation (or SLO) as a direct 
source of turbulence and mixing at the convective boundary.

We begin in \S\,\ref{sec:popI} with a description of
the evolution of relevant characteristics for 
our $3\,M_\odot$, $Z=0.02$ star.
In \S\,\ref{sec:formalism}, we recall the main aspects 
of the formalism
we used to evaluate the impact of IGWs on stellar evolution.
The following sections are devoted
to discussing the pre-main sequence (\S\,\ref{sec:pms}), the main sequence and
the sub-giant branch (\S\,\ref{sec:ms}), the red-giant branch (\S\,\ref{sec:giants}), and 
the clump and the early-AGB (\S\,\ref{sec:AGB}).

\begin{figure}[t]
\centerline{
\psfig{figure=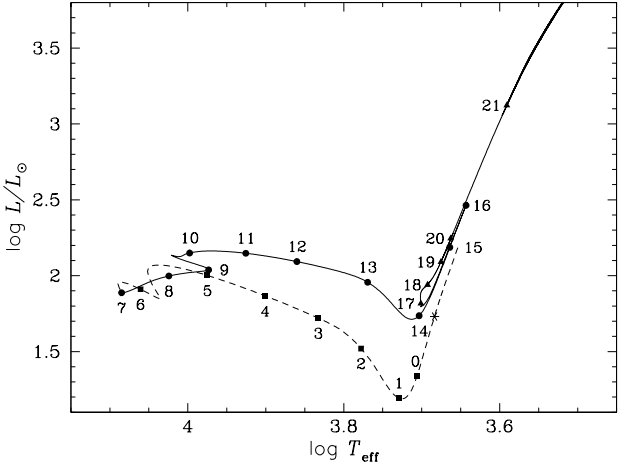,height=6.5cm,angle=-90}
}
\caption{Evolution of our $3\,M_\odot$ Pop\,I model in the HR diagram, 
from the PMS to the early-AGB. Numbers correspond to 
selected evolutionary points that are listed in 
Table~\ref{tab:popI}. For clarity, the PMS is drawn with a dashed line, and 
the corresponding models are identified by squares, while the star symbol corresponds to
the appearance of the convective core.
The circles correspond to evolutionary points on the main sequence, sub-giant, and first 
ascent giant branches, while triangles refer to evolutionary points on the clump and 
on the early-AGB. 
\label{fig:HRI}}
\end{figure}

\begin{figure}[t]
\centerline{
\psfig{figure=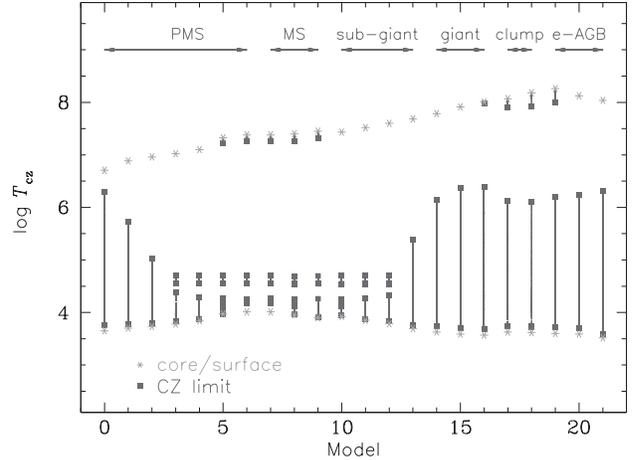,height=6.5cm,angle=-90}
}
\caption{Evolution of the temperature at the boundaries of convection zones for 
the evolutionary points selected for the $3\,M_\odot$ Pop\,I star and given 
in Table~\ref{tab:popI}. Core and surface (i.e., $T_{\rm eff}$) temperatures are also shown.
The evolutionary phases are indicated.
\label{fig:zc3I}}
\end{figure}

\begin{figure*}[t]
\centerline{
\psfig{figure=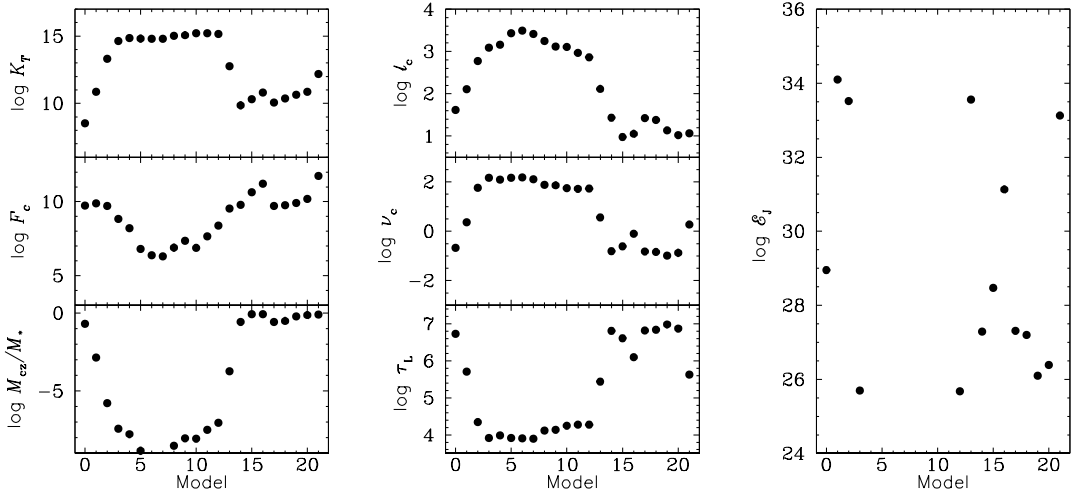,height=7.5cm,angle=-90}
}
\caption{Evolution of the surface convection zone properties for the 
$3\,M_\odot$ Pop\,I 
evolutionary points from Table~\ref{tab:popI}. It shows the thermal conductivity 
$K_T$ (in ${\rm cm^2\,s^{-1}}$) at the convective boundary, 
the convective flux $F_c=\rho v_c^3$ (in ${\rm g\,cm^{-3}}$) 
taken $0.5~H_P$ into the convection zone, the mass of the convection zone, 
the spherical harmonic degree corresponding to the convective scale $\ell_c$, 
the convective frequency $\nu_c$ (in $\mu$Hz), the typical convective timescale 
$\tau_L=\alpha H_P/v_c$, as well as the total energy 
luminosity (in ${\rm g\,cm^2\,s^{-3}}$) prior to any damping in waves 
(this is for waves with $\nu < 3.5\,\mu {\rm Hz}$ and
$1<\ell<60$).
\label{fig:zc2I}}
\end{figure*}

\section{Evolution of relevant characteristics \label{sec:popI}}

In this paper we focus on a $3\,M_\odot$ with a metallicity $Z=0.02$ computed with the 
stellar evolution code STAREVOL (V2.30; Siess et al.~2000). 
Details regarding the code and the input physics may be found in TC05 
and Palacios et al.~(2006).
The model was computed with classical assumptions, \ie, with neither atomic 
diffusion nor rotation.
Convection is treated using the standard mixing length theory with 
$\alpha_{\rm MLT} = 1.75$.
The treatment of convection has an impact on the excited wave spectrum
(see \S\,\ref{sec:spectrum}).

When a star evolves, the characteristics that are relevant to wave excitation
and to momentum extraction by IGWs change. 
We thus chose a series of models (\ie, evolutionary points) to correctly represent 
the evolution of the stellar structure in terms of convection zone properties, 
therefore in terms of IGW characteristics.
The position of the selected points along the evolutionary track in the HR diagram 
for the $3\,M_\odot$ Pop\,I star is shown in Fig.~\ref{fig:HRI}. 

Figure~\ref{fig:zc3I} shows the 
temperature at the boundary of both central and external convection zones at the selected 
evolutionary points, 
which is a way to characterize the depth and extension of these regions as
the surface and central temperatures (also shown in the figure) of the star evolve.
In such an intermediate-mass star, both central hydrogen- and helium-burning occur 
in a convective core. Wave excitation by core convection on the main sequence 
is discussed elsewhere (Pantillon et al.~2007, Paper III). 
Here we only focus on wave excitation by the surface convection zone.

\begin{figure}
\centerline{
\psfig{figure=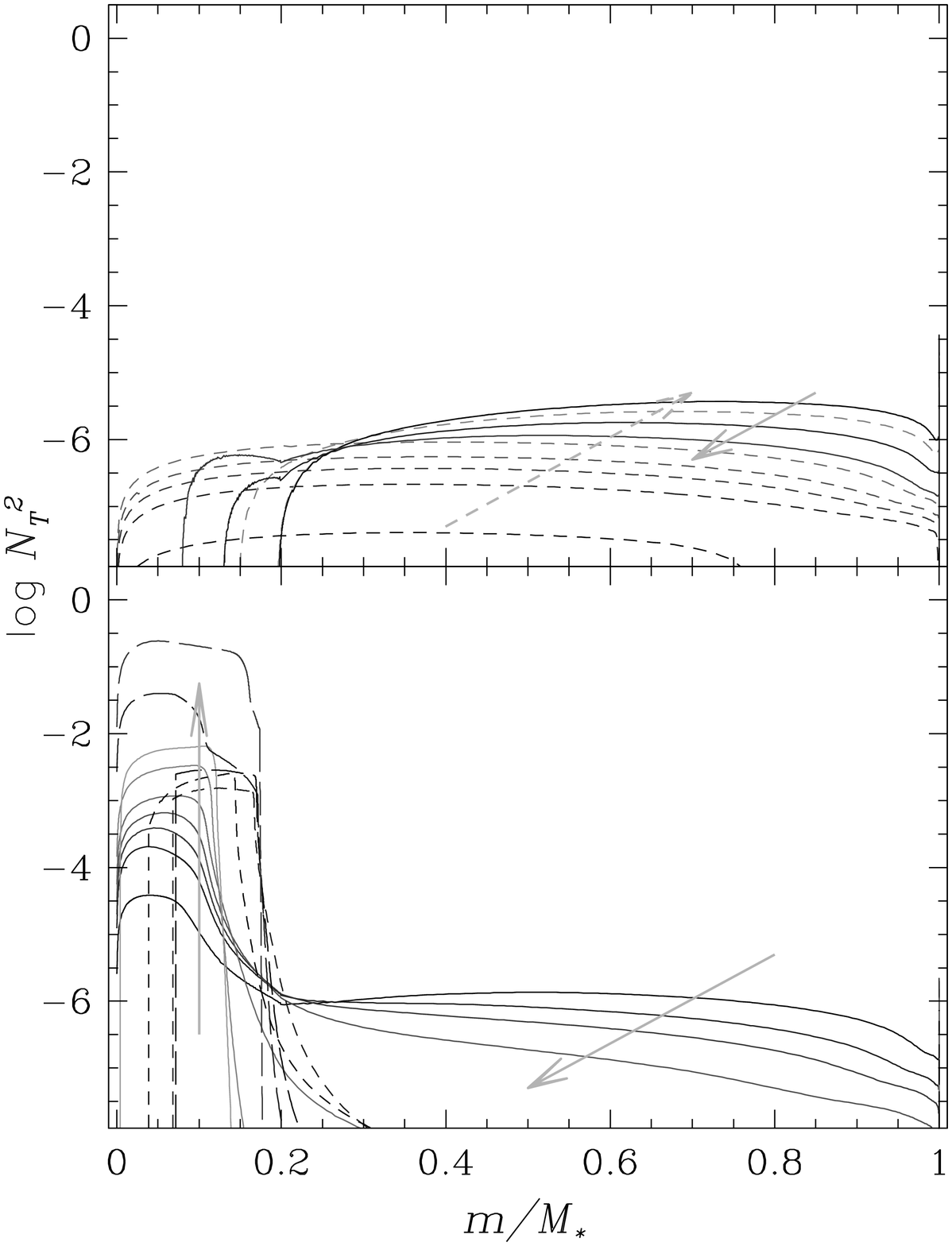,height=10.5cm}
}
\caption{Evolution of the (thermal part of the) Brunt-V\"ais\"al\"a frequency vs mass
for our $3\,M_\odot$ Pop\,I star.
(Top) The dashed lines correspond to PMS models and 
the continuous lines,
to MS models. In PMS models, ${N_T}^2$ increases with evolution, while it decreases 
in MS models. The shade of gray becomes lighter for more evolved models.
The dashed arrow corresponds to the evolution during the PMS and the continuous one 
to the evolution during the MS.
(Bottom) The continuous lines correspond 
to sub-giants and giants, the short-dashed lines to 
the clump, and the long-dashed lines to early-AGB models. 
In the outer layers, ${N_T}^2$ diminishes
with time while it is the opposite in the core.
The arrows indicate time evolution.
\label{fig:BVI}
}
\end{figure}

Figures~\ref{fig:zc2I} and \ref{fig:BVI} illustrate the evolution of the main
properties that are required to understand the behavior of IGWs in the 
framework described in \S\,3. 
Wave excitation is stronger when the convective scale $\ell_c$
is larger. However, when the turn-over timescale $\tau_L$ becomes too large, 
this efficiency diminishes. 
As we shall see in \S~3 and subsequent sections, the combination of these 
two factors produces large differences in the overall efficiency of wave generation 
as the stellar structure evolves.
Another important property for wave-induced transport is the thermal diffusivity 
$K_T$ at the top of the radiative region. 
On the main sequence for the star we focus on, $K_T$ is so large that all 
the low frequency waves
(here, we mean waves with $\sigma < 3.5\,\mu{\rm Hz}$) are dissipated as soon as they are
formed, hence the absence of a filtered flux in the corresponding models (see 
\S~5). 

Finally, damping also depends strongly on the Brunt-V\"ais\"al\"a frequency
(Eq.~\ref{optdepth}), whose thermal part is given by
\beq
{N_T}^2 = \frac{g}{H_P} \delta \lp \nabla_{\rm ad}-\nabla \rp 
\label{BVfT}
\eeq
with $\delta = -(\partial \ln \rho/\partial \ln T)_{P,\mu}$, and other
symbols have their usual meaning. From Fig.~\ref{fig:BVI} we see that  
during the PMS, the star's contraction leads to an increase in gravity and a
reduction of the pressure scale height, which produce a large increase in
${N_T}^2$ when combined. Later, during the giant phase, the expansion 
of the outer layers produces a
large decrease in ${N_T}^2$ outside the hydrogen-burning shell. More important, core
contraction produces a major increase inside this shell; in fact, ${N_T}^2$ becomes so
large there that (low-frequency) waves will be unable to cross that boundary.
A large angular momentum deposition is expected there, provided there is differential
rotation above to produce a bias in the local wave flux.

\section{Formalism \label{sec:formalism}}

The formalism we use to describe IGW properties is extensively described elsewhere
(Papers~I and II, and TC05). 
Here, we only recall the main features of our model and discuss the 
critical physical principles.

\subsection{Wave spectrum \label{sec:spectrum}}

In terms of angular momentum evolution, the relevant parameter is the filtered 
angular momentum luminosity slightly below the convection envelope 
(hereafter CE).
To get that luminosity, we first need to obtain the spectrum of excited waves.
As we did in previous studies, we apply 
the Goldreich, Murray \& Kumar~(1994) formalism to IGWs to calculate this spectrum. 
The energy flux per unit frequency ${\cal F}_E$ is then

\begin{figure*}
\centerline{
\psfig{figure=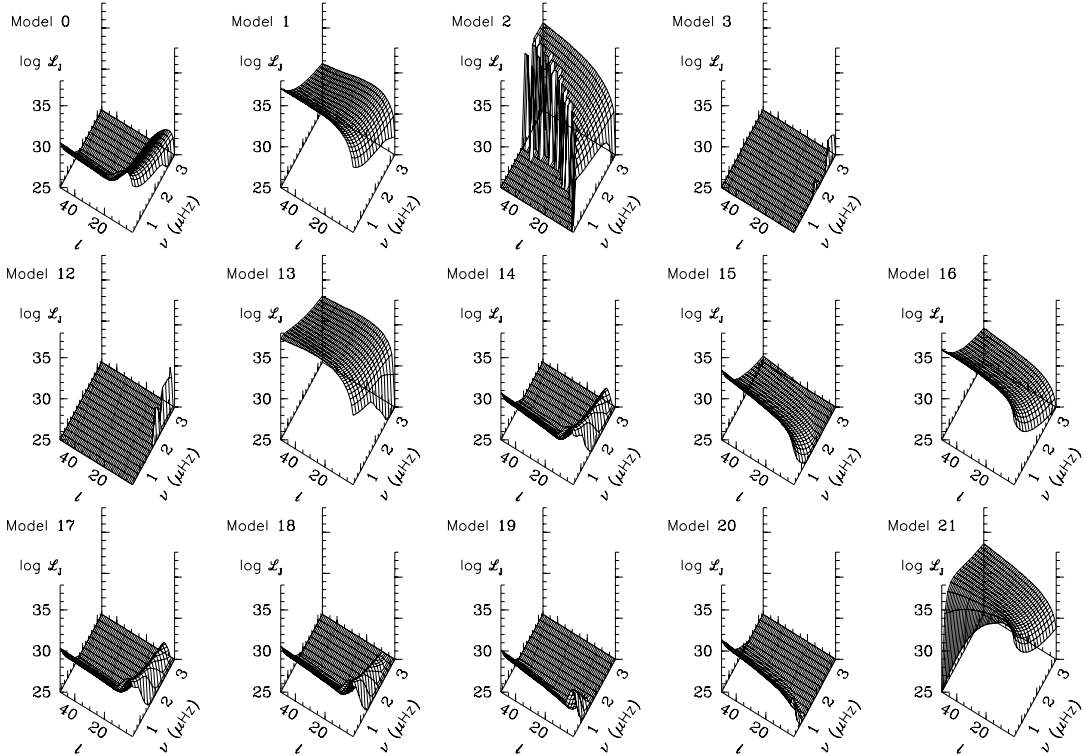,height=12cm,angle=-90}
\vspace*{-1.1cm} }
\caption{Evolution of the spectrum of angular momentum luminosity 
integrated over $\delta \nu=0.2\,\mu{\rm Hz}$
(in ${\rm g\,s^{-3}}$ prior to any damping of IGWs
on the PMS (upper row)
and from the giant branch to the early AGB in the $3\,M_\odot$ Pop\,I star.
Models 4 to 11 have no waves with $\nu < 3.5\,\mu{\rm Hz}$ 
and are thus not shown here.
\label{fig:spectreI}}
\end{figure*}

\begin{eqnarray}
{\cal F}_E \lp \ell, \omega \rp &=& \frac{\omega^2}{4\pi} \int dr\; \frac{\rho^2}{r^2}
   \left[\left(\frac{\partial \xi_r}{\partial r}\right)^2 +
   \ell(\ell+1)\left(\frac{\partial \xi_h}{\partial r}\right)^2 \right]  \nonumber \\
 && \times  \exp\left[ -h_\omega^2 \ell(\ell+1)/2r^2\right] \frac{v_c^3 L^4 }{1
  + (\omega \tau_L)^{15/2}},
\label{gold}
\end{eqnarray}
where
$\xi_r$ and $[\ell(\ell+1)]^{1/2}\xi_h$ are the radial and horizontal
displacement wave functions, which are normalized to unit energy flux just
below the convection zone, $v_c$ is the convective velocity, 
$L=\alpha_{\rm MLT} H_P$ the radial
size of an energy bearing turbulent eddy, $\tau_L \approx L/v_c$ the
characteristic convective time, and $h_\omega$ is the
radial size of the largest eddy at depth $r$ with characteristic frequency of
$\omega$ or higher ($h_\omega = L \min\{1, (2\omega\tau_L)^{-3/2}\}$).
The radial wave number $k_r$ is related to the horizontal wave number $k_h$ by
\beq
k_r^2 = \lp \frac{N^2}{\sigma^2} -1 \rp k_h^2 = 
\lp \frac{N^2}{\sigma^2} -1 \rp \frac{\ell \lp \ell +1 \rp}{r^2} \label{kradial}
\eeq
where $N^2$ is the Brunt-V\"ais\"al\"a frequency. In the convection zone, the mode is
evanescent and the penetration depth varies as $\sqrt{\ell \lp \ell +1 \rp}$.
The momentum flux per unit frequency ${\cal F}_J$ is then related to the 
kinetic energy flux by  
\begin{equation}                                                                          
{{\cal F}_J}\lp m, \ell, \omega \rp = \frac{2m}{\omega} {\cal F}_E\lp \ell, \omega \rp  
\end{equation}   
(Goldreich \& Nicholson 1989; Zahn, Talon \& Matias 1997). 
We integrate this quantity horizontally to get an angular momentum luminosity
\beq
{\cal L}_J = 4 \pi r^2 {\cal F}_J
\eeq
which, in the absence of dissipation, has the advantage of being a conserved quantity 
(Bretherton~1969; Zahn et al.~1997).
This angular momentum luminosity, integrated over $\delta \nu=0.2\,\mu{\rm Hz}$,  
is illustrated in Fig.~\ref{fig:spectreI} for our selected evolutionary 
points, that have low frequency waves with $\nu < 3.5\,\mu{\rm Hz}$.

Each wave then travels inward and is damped by thermal diffusivity and by viscosity.
The local momentum luminosity of waves is given by
\beq
{\cal L}_J(r) = \sum_{\sigma, \ell, m} {{\cal L}_J}_{\ell, m} \lp r_{\rm cz}\rp
\exp \lc -\tau(r, \sigma, \ell) \rc \label{locmomlum}
\eeq
where `${\rm cz}$' refers to the base of the convection zone.
$\tau$ corresponds to the integration of the local damping rate and takes into account 
the mean molecular weight stratification
\beq
\tau(r, \sigma, \ell) = \lc \ell(\ell+1) \rc ^{3\over2} \int_r^{r_c} 
\lp K_T + \nu_v \rp \; {N {N_T}^2 \over
\sigma^4}  \left({N^2 \over N^2 - \sigma^2}\right)^{1 \over 2} {\diff r
\over r^3} \label{optdepth}
\eeq
(Zahn et al. 1997).
In this expression, ${N_T}^2$ is the thermal part of the Brunt-V\"ais\"al\"a 
frequency (see Eq.\ref{BVfT}), 
$K_T$ the thermal diffusivity, and $\nu_v$ the (vertical) turbulent viscosity.
Finally $\sigma$ is the local, Doppler-shifted frequency
\beq
\sigma(r) = \omega - m
\lc \Omega(r)-\Omega _{\rm cz} \rc \label{sigma}
\eeq
and $\omega$ is the wave frequency in the reference frame of the convection
zone. 
Let us mention that only the radial velocity gradients are taken into account 
in this expression for damping. This is because angular momentum transport
is dominated by the low-frequency waves ($\sigma \ll N$), which implies
that horizontal gradients are much smaller than vertical ones (\cf Eq.~\ref{kradial}).

When meridional circulation, turbulence, and waves are all taken into account, 
the evolution of angular momentum follows
\beqan
\rho \dtt \lc r^2 {\Omega}\rc &= &
\frac{1}{5 r^2} \drr \lc \rho r^4 \Omega U \rc 
+ \frac{1}{ r^2} \drr \lc \rho \nu_v r^4 \dr{\Omega} \rc 
\nonumber \\
&& - \frac{3}{8\pi} \frac{1}{r^2} \drr{{\cal L}_J(r)}
\label{ev_omega}
\eeqan
(Talon \& Zahn 1998),  
where $U$ is the radial meridional circulation velocity.
Horizontal averaging was performed for this equation, and meridional circulation was
considered only at first order. In this paper, this equation is used only to calculate the
fast SLO's dynamics (see \S\,\ref{sec:SLO}) and thus, $U$ is neglected. The 
complete equation is used when longer timescales are involved as discussed \eg\, in
Charbonnel \& Talon~(2005).

In this work, we neglect any contribution from convective overshoot due to the lack
of a usable prescription (although work is underway to 
include this effect, Belkacem et al., in preparation). 
We expect this contribution to be more efficient for low-frequency, low-degree 
waves, and as such, it would have an impact mostly on the timescale for
angular momentum extraction in the core (see \S\,\ref{sec:secular}).

\begin{figure*}
\centerline{
\psfig{figure=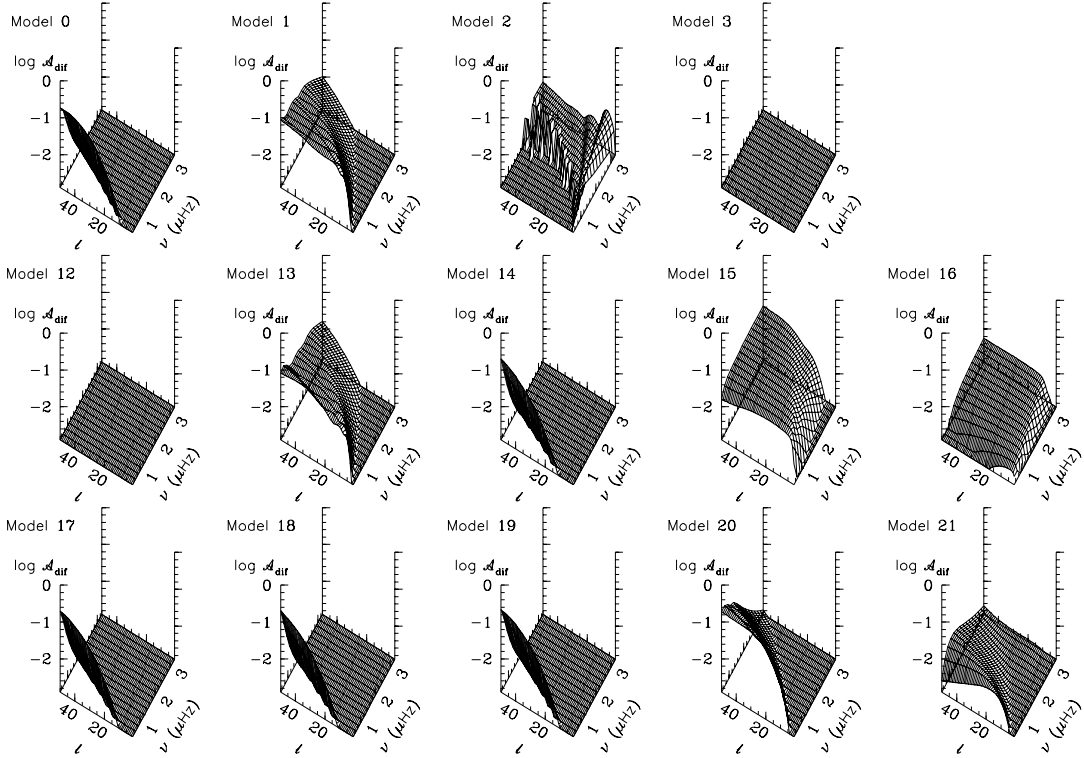,height=12cm,angle=-90}
\vspace*{-1.1cm} }
\caption{Differential wave amplitude variation 
over a depth of $0.05\,R_*$ (see text).
\label{fig:specdepI}}
\end{figure*}

\subsection{Shear layer oscillation (SLO) \label{sec:SLO}}
One key feature when looking at the wave-mean flow interaction is that the dissipation of
IGWs produces an increase in the local differential rotation, which is caused by the
increased dissipation of waves that travel in the direction of the shear (see
Eqs.~\ref{optdepth} and \ref{sigma}). In conjunction with
viscosity, this leads to the formation of an oscillating doubled-peak shear layer that
oscillates on a short timescale (Gough \& McIntyre~1998; Ringot~1998; 
Kumar, Talon \& Zahn~1999). 
This shear layer oscillation (hereafter SLO) occurs if the deposition of angular
momentum by IGWs is large enough when compared with viscosity\footnote{When the wave 
flux is too small, one simply reaches a stationary regime where the angular momentum 
deposition by IGWs is compensated by viscous diffusivity.} (Kim \& MacGregor~2001; 
see also a discussion in TC05, \S~3.1.). 

To calculate the turbulence associated with this oscillation, we relied on
a standard prescription for shear turbulence away from regions
with mean molecular weight gradients
\beq
\nu_v = \frac{8}{5} Ri_{\rm crit} K_T \frac{\lp r \diff \Omega/\diff r \rp ^2}
{{N_T}^2}, 
\label{maeder95}
\eeq
which takes radiative losses into account (Townsend~1958;
Maeder~1995). This coefficient is time-averaged over a complete
oscillation cycle (for details, see TC05, \S\,3.1).

The first feature we are interested in is whether or not the formation of
such an SLO is possible at various evolutionary stages, since the associated turbulence
could produce mixing in a small region below the surface convection zone.
Figure\,\ref{fig:specdepI} shows the differential wave amplitude 
variation in the region of SLO formation. The plotted quantity corresponds to
\beq
{\cal A}_{\rm dif}=\sum_{r_{\rm cz}}^{r_{\rm cz}+0.05\,R_*} \left| \Delta A_{\rm prograde}-
\Delta A_{\rm retrograde} \right|
\eeq
where $\Delta A$ corresponds to the local amplitude variation,
caused by radiative damping (see Eq.~\ref{optdepth}) 
for the $\ell=m$ wave and for a local gradient of $0.001\,\mu{\rm
Hz}/0.05\,R_*$.
This choice of differential rotation is quite arbitrary, but TC05 (see their \S\,3.2)
show that, as long as
differential rotation is not too strong (that is, it remains below about 
$1\,\mu{\rm Hz}/0.05\,R_*$), differential damping is linear. The value
of $0.05\,R_*$ is typical of the thickness of the SLO (TC05).\\
When ${\cal A}_{\rm dif}$ is large, prograde and retrograde waves
are deposited at different locations in the region of SLO formation and thus
contribute to the potential generation of an SLO. As the wave frequency rises, 
the Doppler shift is felt less by the star and ${\cal A}_{\rm dif}$ remains small. 
For frequencies that are too low, the
waves can be damped even before the Doppler shift may take place (see e.g.
Model\,1). 

\begin{figure}
\centerline{
\psfig{figure=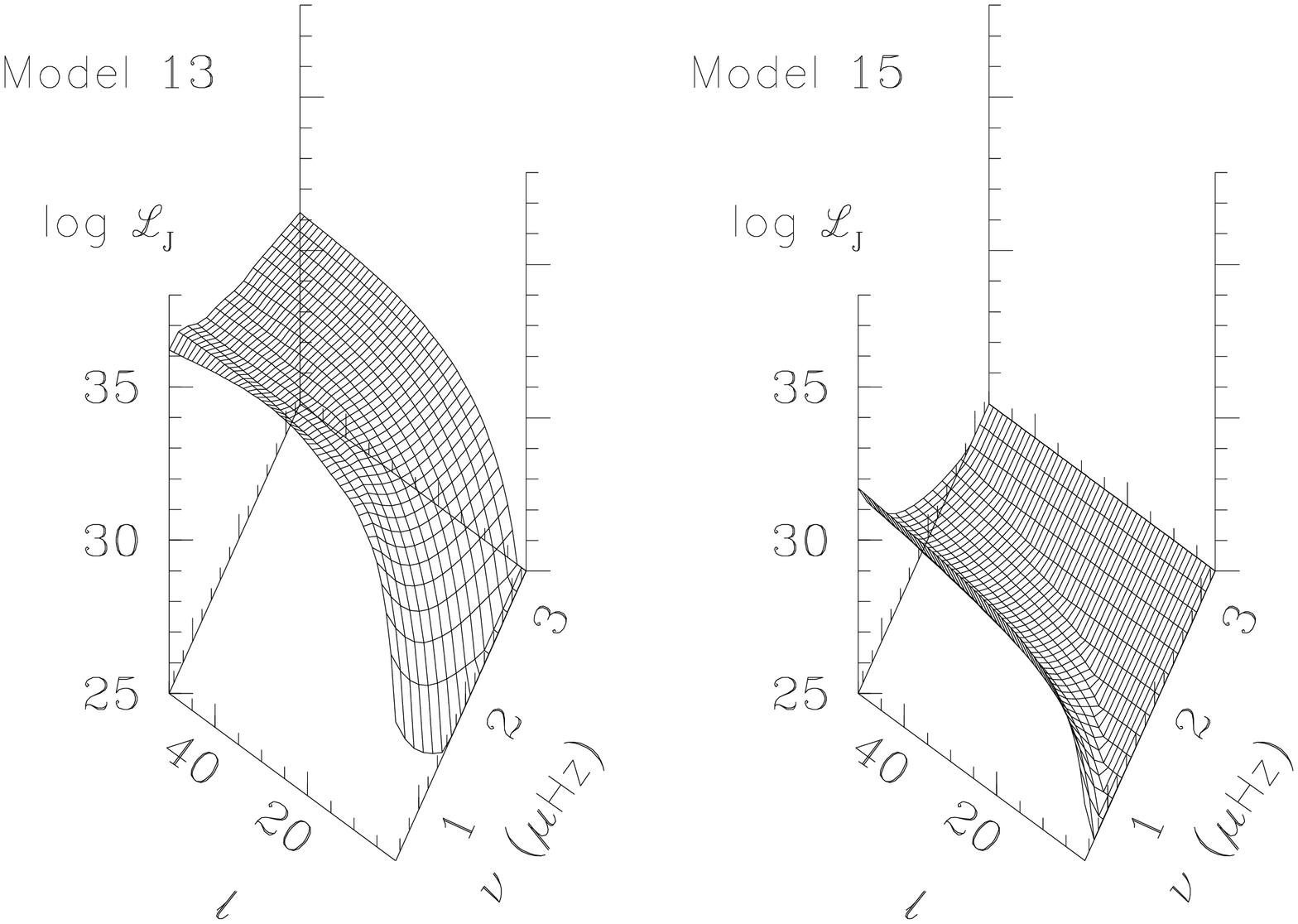,height=5cm,angle=-90}
\vspace*{-0.2cm} }
\caption{Differential wave angular momentum luminosity variation 
over a depth of $0.05\,R_*$ for models 13 and 15.
\label{fig:spec1315}}
\end{figure}

To discuss the formation of the SLO, ${\cal A}_{\rm dif}$ should be
multiplied by the angular momentum luminosity of waves, which is shown in
Fig.\,\ref{fig:spectreI} for the GMK model. 
This was done for example for models 13 and 15 in Fig.\,\ref{fig:spec1315}.
For these two models, the waves that would give rise to the appearance of an SLO are
quite similar, as the differential amplitude figure shows (Fig.\,\ref{fig:specdepI}).
However, there is an SLO only in model 13, because in model 15, the wave angular momentum
luminosity corresponding to such waves is simply too small.
We note that, for any other 
excitation model, an SLO would be obtained for an angular momentum luminosity on the order of 
$10^{33}-10^{34}\,$ (or more) in the region with ${\cal A}_{\rm dif}\simeq 0.01$ to $0.1$.
Therefore, for any excitation model we can rapidly evaluate whether an SLO should 
develop from our Fig.\ref{fig:specdepI}. 

The existence of such an SLO has been challenged by 
Rogers \& Glatzmeier~(2006, hereafter RG06).
In 2--D numerical simulations of penetrative convection in the Sun, they 
find that IGWs are efficiently excited by convection, as did the other authors who studied
this (Hurlburt et al.~1986,~1994; 
Andersen~1994; Nordlund et al.~1996; Kiraga et al.~2003; Dintrans et al.~2005; Rogers \& 
Glatzmeier~2005a,b). 
In these simulations, wave spectra are much broader than the ones
predicted in theoretical models based on Reynolds stresses (Kumar, Talon, \& Zahn~1999,
hereafter KTZ99).
However, contrary to the expectations of other authors (see in particular the
discussion by Kiraga et al.~2003), RG06
obtain an energy flux\footnote{In their simulation, which has a lower Reynolds number
than the Sun, but a larger convective flux, the total wave flux is rather large, 
and it scales as
\beq
F_{\rm waves} = F_T \frac{v}{c_s}
\eeq
with $F_T$ the total flux, $v$ the velocity of turbulent eddies and $c_s$ the
sound velocity.} that is smaller 
{\it in the low-frequency regime} 
than the one calculated 
by KTZ99, and that, even though one expects the excitation by convective plumes that has been
ignored by KTZ99 to yield a large contribution to wave generation.

\begin{figure*}
\centerline{
\psfig{figure=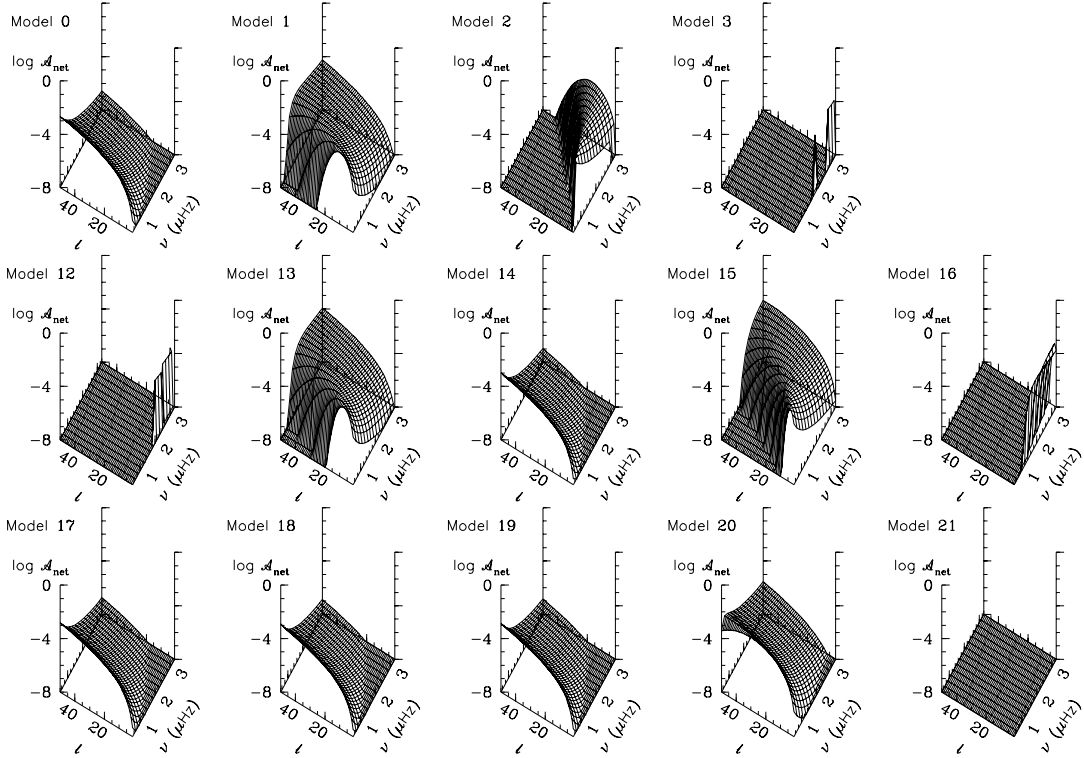,height=12cm,angle=-90}
\vspace*{-1.1cm} }
\caption{Net wave amplitude beyond a depth of $0.05\,R_*$ (see text).
\label{fig:specnetI}}
\end{figure*}

In their simulations, RG06 thus have a much weaker wave flux 
in the low-frequency regime than what
is required to generate an SLO (see the above discussion on 
Fig.~\ref{fig:spec1315}),
although the one for the high frequency waves is much larger.
It may thus not be surprising that RG06 do not find such an SLO in their simulation.
Furthermore, the size of the SLO being smaller (in KTZ99) than the difference between the
rotation rate at the base of the convection zone (at the equator) and the base of the
tachocline, an SLO would not appear as a shear reversal in their simulation.
Last, the RG06 simulation is simply not long enough to show the SLO. 
As pointed out by
the authors, in their numerical simulation there is differential rotation in the tachocline 
driven by convective plumes. But the total duration of the simulation being only one year, it
would be impossible to see a superposed SLO since its timescale is expected to
be of about a few years, according to the KTZ99 calculations. 
Furthermore, since RG06 have a reduced wave flux compared to KTZ99, this timescale
would be even longer, and thus clearly beyond the possibilities of the RG06 simulation.

\begin{table*}
\caption{Selected models along the evolutionary track of a $3\,M_\odot$ Pop\,I star.
The filtered angular momentum luminosity ${{\cal L}_J}^{\rm fil}$ is taken as 
$0.05\,R$ below the CZ (or at $R_{\rm cz}/2$ if it is larger)
for a differential rotation of $\delta \Omega = 0.001\,\mu{\rm Hz}$.
Also provided are the star's moment of inertia of the radiative
zone $I_{\rm R}$ and the resulting timescale 
$\tau=I_R \delta \Omega/{\cal L}_J^{\rm fil}$ for IGWs.
The last column gives the time $\Delta t$ 
between consecutive models (with the convention $\Delta t_i = t_{i+1} - t_i$).
No waves are generated in models 4 to 11.
\label{tab:popI}}
\begin{center}
\begin{tabular}{c|c|c|c|c|c|c}
      & Evolutionary&& ${{\cal L}_J}^{\rm fil}$ & $I_{\rm R}$ & $\tau$ & $\Delta t$ \\
Model & Status& SLO  & (${\rm g\,cm^2\,s^{-2}}$) &(${\rm g\,cm^2}$)&(${\rm yr}$) & (${\rm yr}$)\\
\hline
\hline
0 & PMS       & no & $-3.4 \times 10^{32}$& $7.7\times 10^{55} $ & $4.5 \times 10^7$ & $7.2\times 10^5$  \\
1 &           & yes & $-2.5 \times 10^{37}$& $7.4\times 10^{55} $ & $5.8 \times 10^3$ & $7.7\times 10^5$  \\
2 &           & yes & $-1.5 \times 10^{35}$& $5.5\times 10^{55}$ & $7.1 \times 10^4$ & $3.1\times 10^5$  \\
3 &           & no & $-9.2 \times 10^{24}$& $4.2\times 10^{55}$ & $9.2 \times 10^{14}$ & $2.0\times 10^5$  \\
4 &           & no & $ -$& $ - $ & $-$& $2.0\times 10^5$  \\
5 &           & no & $ -$& $ - $ & $-$& $8.8\times 10^5$  \\
6 &           & no & $ -$& $ - $ & $-$& $8.8\times 10^6$ \\
\hline
7 & MS        & no & $-$& $ - $ & $-$ & $2.2\times 10^8$  \\
8 &           & no & $-$& $ - $ & $-$ & $7.9\times 10^7$  \\
9 &           & no & $-$& $ - $ & $-$ & $1.3\times 10^7$  \\
\hline
10 & sub-giant & no & $ -$& $ - $ & $-$& $7.9\times 10^6$  \\
11 &           & no & $ -$& $ - $ & $-$& $1.4\times 10^6$  \\
12 &           & no & $ -9.5 \times 10^{26}$& $ 5.2 \times 10^{55}  $ & $1.1\times 10^{13}$ &
$9.6\times 10^5$ \\
13 &           & yes & $ -5.5 \times 10^{36}$& $ 1.0 \times 10^{56}  $ & $3.8\times 10^{3}$ &
$9.8\times 10^5$ \\
\hline
14 & giant     & no & $ -1.9 \times 10^{31}$& $ 1.6 \times 10^{56}  $ & $1.7\times 10^{9}$ &
$1.9\times 10^6$ \\
15 &           & no & $ -4.9 \times 10^{27}$& $ 1.4 \times 10^{54}  $ & $5.6\times 10^{10}$ &
$1.5\times 10^6$ \\
16&           & yes & $ -1.6 \times 10^{28}$& $ 1.8 \times 10^{53}  $ & $2.2\times 10^{9}$ &
$4.6\times 10^7$ \\
\hline
17& clump       & no & $ -8.0 \times 10^{30}$& $ 1.8 \times 10^{56}  $ & $4.4\times 10^{9}$ &
$9.4\times 10^7$ \\
18&           & no & $ -1.3 \times 10^{31}$& $ 1.9 \times 10^{56}  $ & $2.9\times 10^{9}$ &
$9.8\times 10^6$ \\
\hline
19& early AGB       & no & $ -4.7 \times 10^{30}$& $ 5.9 \times 10^{55}  $ & $2.5\times 10^{9}$&
$1.1\times 10^7$  \\
20&           & no & $ -2.7 \times 10^{30}$& $ 1.2 \times 10^{55}  $ & $9.0\times 10^{8}$ &
$8.0\times 10^6$ \\
21&           & yes& $ -1.0 \times 10^{29}$& $ 2.4 \times 10^{52}  $ & $4.7\times 10^{7}$ \\
\hline
\end{tabular}
\end{center}
\end{table*}

\subsection{Filtered angular momentum luminosity \label{sec:secular}}
We must also look at the secular effect of IGWs in the deep interior 
(see TC05 for details).
In the presence of differential rotation, the dissipation of prograde and retrograde waves 
in the SLO is
not symmetric, and this leads to a finite amount of angular momentum being deposited in the
interior beyond the SLO.
Figure\,\ref{fig:specnetI} shows the net amplitude of waves
at a depth of $0.05\,R_*$, defined by
\beq
{\cal A}_{\rm net}=A_{\rm retrograde}-A_{\rm prograde}
\eeq
for a local gradient of $0.001\,\mu{\rm Hz}/0.05\,R_*$. Multiplied
by the angular momentum luminosity of waves, this 
is the filtered angular momentum luminosity ${{\cal L}_J}^{\rm fil}$ given 
in Table~\ref{tab:popI} for our selected evolutionary points.
Let us mention that, in fact, the existence of an SLO is not even required
to obtain this differential damping between prograde and retrograde waves, so 
as long as differential rotation exists at the base of the convection zone, waves will 
have a net impact on the rotation rate of the interior.
The timescale required to delete the imposed differential rotation is
of the order of
\beq
\tau = I \delta \Omega / {{\cal L}_J}^{\rm fil}.
\label{eq:tau}
\eeq

The SLO's dynamics is studied by solving Eq.~(\ref{ev_omega}) with small timesteps
and using the whole wave spectrum, while for the secular evolution of the star,
one has to instead use the filtered angular momentum luminosity
(see TC05 for details).
All the relevant quantities are given 
in Table~\ref{tab:popI} for our selected evolutionary points.
Let us also mention here that differential damping is required both for the SLO and for the
filtered angular momentum luminosity. 
Since this relies on the Doppler shift of the frequency (see Eqs.~\ref{optdepth} 
and \ref{sigma}), angular momentum redistribution will be dominated by
the low-frequency waves, which experience a larger Doppler shift, but that
is not so low that they will be immediately damped. Numerical
tests indicate that this occurs around ($\sigma \simeq 1~\mu {\rm Hz}$).

\begin{figure*}
\centerline{
\psfig{figure=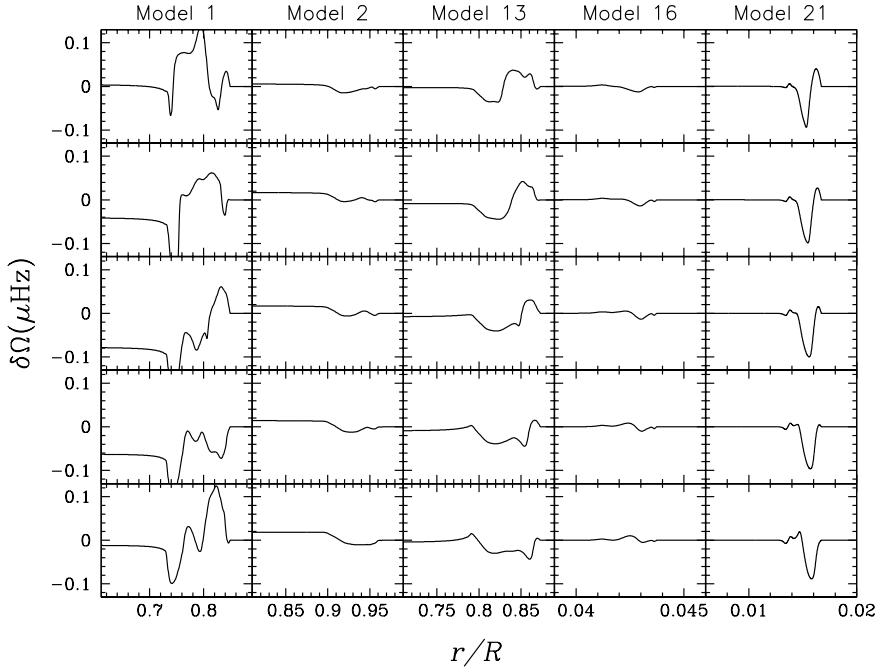,height=9cm,angle=-90}}
\caption{Shear layer oscillation in
the $3~M_{\odot}$ Pop\,I model at the evolutionary points where 
this feature appears.
For each evolutionary point, successive profiles are separated by $\Delta
t=5\,{\rm yr}$.
\label{fig:slo}}
\end{figure*}

\section{Pre-main sequence \label{sec:pms}}

We start with a fully convective contracting star. 
As it descends along the Hayashi track, a radiative core appears (Fig.~\ref{fig:HRI}). 
In mass coordinates, the top of this radiative region migrates towards the surface until 
the star reaches the main sequence. This is accompanied by a growth of the characteristic 
convective length scale at the bottom of the convection zone 
$\ell_c=2 \pi r_{\rm cz} / \alpha H_P$ (see Fig.~\ref{fig:zc2I}), which 
temporarily favors the generation of waves (see models 0 to 2). 

However, with further retraction of the envelope, the thermal diffusivity $K_T$ 
at the convective boundary increases,
which favors the disappearance of low-frequency and/or large degree waves 
(see Fig.~\ref{fig:spectreI} and Eq.~\ref{optdepth}). 
This explains the reduction of the total energy flux 
at the end of the PMS (models 3 to 6) and on the MS.

\begin{figure}
\centerline{
\psfig{figure=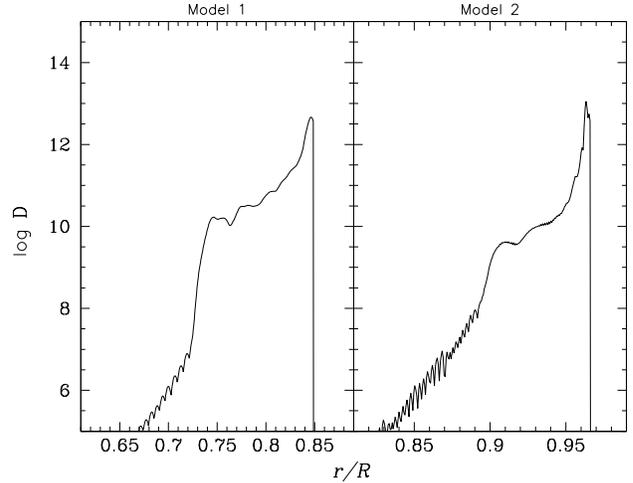,height=6.5cm,angle=-90}
}
\caption{Turbulent wave-induced diffusion coefficient at the convection zone boundary
on the pre-main sequence in the $3\,M_\odot$ Pop\,I star
\label{fig:difPMS}}
\end{figure}

During the pre-main sequence, an SLO exists only for a very short period of time.
The evolution of the SLO for models~1 and 2 is shown in Fig.~\ref{fig:slo}, 
where the successive profiles are separated by 5~years.  
The larger amplitude in model~1 compared to model~2 can be understood in terms
of thermal diffusivity: since $K_T$ increases with radius, as the top of the radiative
zone migrates outwards with evolution, waves are dissipated more efficiently in
model~2 than in model~1. In model~3, damping is so large that the formation of
a SLO is no longer possible, and all low-frequency waves are dissipated in model~4. 
The turbulent diffusion coefficient associated with this SLO, $D$ (see Eq.~\ref{maeder95}
and the discussion around it), is shown 
in Fig.~\ref{fig:difPMS} as a function of reduced radius. 
While the shear is much stronger in model~1 than in model~2, the magnitude 
of $D$  and width 
of the turbulent regions are quite similar. This is because the reduced shear
is compensated for by the larger thermal diffusivity in model~2 (\cf\,Eq.~\ref{maeder95}). 
In both models,
$D$ is relatively high just below the convective envelope, and it drops rapidly
with depth. It is negligible in the region with a temperature 
(\ie\,$T \apprge  2.5 \times 10^6\,{\rm K}$) high enough for efficient lithium burning.  
This temperature is attained at reduced radii of $r/R \sim 0.43$ and $0.41$ 
in models~1 and 2 respectively.
This implies that, for this mass, the transport of elements through wave-induced turbulence
has no impact on the pre-main sequence evolution of the surface abundance 
of the light elements 
such as lithium, beryllium, or boron (the latest two elements burning at an even 
higher temperature)\footnote{No LiBeB depletion is expected on the PMS
in a classical model for a Pop\,I star of that mass, that is due to the CE base too 
cool during that phase (see Fig.~\ref{fig:zc3I}).}.

The role of IGWs is more uncertain for the rotation profile. 
In the case of rapid rotation ($v \apprge 100\,{\rm km\,s^{-1}}$), the initial
rotation profile is deleted by meridional circulation in a fraction of the main-sequence 
lifetime. In a slow rotator, however, this is not the case, and the memory 
of angular momentum transport by IGWs will influence 
mixing\footnote{The role of the evolution of the 
rotation profile on main sequence mixing is discussed further in Talon, Richard \&
Michaud (2006), \S\,3.2.}. Detailed evolutionary calculations of angular momentum
in the presence of internal waves should thus take this evolutionary phase into account 
especially in the case of slow rotators (Charbonnel \& Talon, in preparation).

\section{Main sequence and sub-giant branch \label{sec:ms}}

On the main sequence (models~7 to 9) and early on the sub-giant branch (models~10 and 11), 
the surface convection zone of the star remains quite
shallow and contains no more than $\sim 3 \times 10^{-8}\,M_*$. In the bottom part of
the convection zone, which is the main driver of IGWs, the mean convective
flux is too small ($\sim 0.001\,L_\odot$) to excite IGWs efficiently 
(see Fig.~\ref{fig:zc2I}).
Furthermore, the thermal diffusivity below the convection zone is quite large,
and perturbations traveling into the radiative zone
are rapidly damped instead of becoming waves. This is the case in particular
for all the low-frequency waves that are needed for driving an SLO\footnote{We
recall that thermal damping is proportional to $1/\sigma^4$, see Eq.~\ref{optdepth}.} 
(for waves with a frequency $\nu < 3.5~\mu{\rm Hz}$, ${\cal F}_J=0$).
There is thus no SLO or any secular effect on the rotation profile from this
exterior convection zone, both on the MS and at the beginning of the sub-giant branch 
in such an intermediate-mass star.

This result agrees with the conclusions of Paper~I, i.e., 
the total momentum luminosity in waves drops dramatically in main sequence 
Pop\,I stars with initial masses higher than $\sim 1.4\,M_{\odot}$, 
\ie\,stars originating from the left side of the Li dip, compared to stars 
with lower initial mass. 

Angular momentum transport by IGWs becomes important as the star moves farther along
the sub-giant branch.
Indeed the retraction of the convective envelope is accompanied by a diminution 
of the thermal diffusivity and an increase in the convective flux.
To properly assess the importance of waves in the star's rotation, we should
compare the timescale 
$\tau$ (Eq.~\ref{eq:tau}) with the lifetime of various evolutionary stages;
both quantities are given in Table~\ref{tab:popI}.
Model~13 is of special interest here because it lies at the end of the sub-giant branch 
(see Fig.~\ref{fig:HRI}) and supports a particularly large wave flux, 
due to a unique combination
of many factors (including the convective time- and length-scales, see Fig.~\ref{fig:zc2I}).
This implies that possible differential rotation, which could be a relic of the star's 
main sequence history and subsequent contraction, will be strongly reduced by IGWs 
when the star leaves the Hertzsprung gap. This will have a profound impact on the 
subsequent evolution.
Let us also mention that, close to the core, waves could actually create a large differential 
rotation. This was observed in the numerical simulation by RG06 and is 
also visible when one looks at the Fig.~3 of
Talon et al.~(2002), where, close to the core, strong
shears appear in the early stages of the simulation. Detailed calculations, including
meridional circulation, shear turbulence, and wave induced angular momentum transport, are
required to come to any conclusion other than that the wave will have a strong impact at
this stage.

\begin{figure}
\centerline{
\psfig{figure=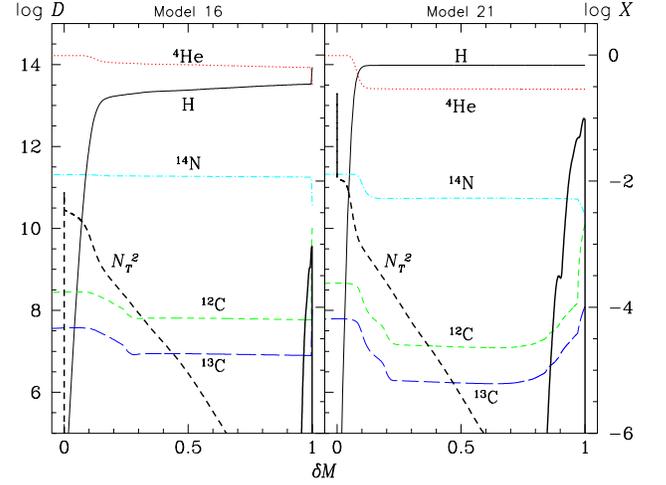,height=6.5cm,angle=-90}
}
\caption{Turbulent wave-induced diffusion coefficient at the convection zone boundary
in the $3\,M_\odot$ Pop\,I star (left) on the red giant branch 
and (right) on the early-AGB 
(wide continuous curve, left scale).
Superimposed is the logarithm of the mass fraction of H, $^4$He, $^{12}$C,
$^{13}$C and $^{14}$N (right scale).
The abscissa is the scaled mass coordinate $\delta M$ which equals
0 at the base of the hydrogen-burning shell and 1 at the base of the CE 
(see text). Also shown is the Brunt-V\"ais\"al\"a frequency (wide dashed curve, right scale).
\label{fig:difgeantes}}
\end{figure}

\section{Red giant branch \label{sec:giants}}

As the star evolves to the red giant branch, the surface convection zone deepens, 
and the thermal conductivity below the convection envelope is reduced
with the photon mean free path (models~12 to 16, see Fig.~\ref{fig:zc2I}). 
Wave excitation during this phase is quite similar to what has been 
observed in PMS stars.

A SLO appears at the tip of the RGB (Model~16). 
The corresponding diffusion coefficient is shown in Fig.~\ref{fig:difgeantes},  
together with the abundance profile of the lightest elements that are affected by shell
hydrogen-burning. These quantities are plotted against $\delta M$, 
which is a relative mass coordinate allowing for a blow-up of the 
radiative region above the hydrogen-burning shell (hereafter HBS) 
and is defined as
\begin{equation}
\delta M = \frac{M_r - M_{\rm HBS}}{M_{\rm BCE} - M_{\rm HBS}}.
\end{equation} 
Here $\delta M$ is equal to 1 at the base of the convective envelope and 0 at
the base of the HBS (where $X = 10^{-7}$).
The abundance profiles at that phase have been modified by dilution 
during the first dredge-up episode and are being reconstructed by shell hydrogen-burning. 
We see that $D$ drops very rapidly just below the CE. Additionally, 
the star leaves that phase very rapidly. As a consequence, 
wave-induced turbulence is not expected to modify the surface abundances
on the RGB for such an intermediate-mass Pop\,I star. 
 
The rotation profile should hardly be influenced by IGWs for that star
on the RGB, in view of the timescale $\tau$ (Eq.~\ref{eq:tau}) which 
is now almost two orders of magnitude higher than the lifetime at that 
stage (see Table~\ref{tab:popI}).

\section{Clump and early-AGB\label{sec:AGB}}
Model~16 marks the ignition of central He-burning in the non-degenerate core, 
and model~17 corresponds to the arrival of the star in the so-called clump 
in the HR diagram. 
At that moment the convective envelope strongly retreats in mass, 
before deepening again slowly until the central exhaustion of helium, 
which occurs for model~18\footnote{The variations in the depth of the CE 
at that phase are hardly seen in Fig.~\ref{fig:zc2I}, because the total
variations represented for $\log\,(M_{\rm cz}/M_*)$ cover nine orders of magnitude.
However, the value for $\log\,(M_{\rm cz}/M_*)$ in models~16, 17, and 21 is
-0.854, -0.138, and -0.747, respectively.}. 
Then the star moves to the early-AGB. 
During that period, the total energy flux in waves remains relatively modest
and there is no SLO. 

The situation changes drastically in model~21 however.
At that point, the convective envelope has reached its maximum depth,  
which stays almost constant until the first thermal pulse occurs 
(at $\Delta t \approx 4\times 10^5\,{\rm yr}$ after model~21). 
During that second dredge-up, the convective envelope has engulfed most (but not all) 
of the more external step of the $^{14}$N profile left behind by the CN-cycle, which explains 
the steep gradients in the abundance profiles of that element and of the carbon isotopes 
just below the convective envelope. In this region, $D$ is relatively high, and 
we expect that the associated mixing leads to an additional slight variation in the surface 
abundance of $^{12}$C and $^{13}$C (but not of the carbon isotopic ratio since CN is 
at the equilibrium), and of $^{14}$N with respect to standard second dredge-up.

For the transport of angular momentum, the rapid increase in the Brunt-V\"ais\"al\"a
frequency in the HBS could produce strong shears that, in complete calculations taking 
rotational mixing into account, would be an important source of mixing during the early-AGB.
This would modify the composition close to the core and would strongly modify 
the subsequent evolution.

\section{Conclusion}

In this paper, we examined when and how internal gravity waves 
generated by surface convection zones are expected to have
an impact on stellar models that only have shallow surface convection zones
during the main sequence. We find that several evolutionary stages can
experience angular momentum redistribution by IGWs:
\begin{itemize}
\item During the pre-main sequence, just as the star leaves the Hayashi track,
waves are efficiently excited and filtered. While this has no impact on 
light element abundances and only has a limited effect on fast rotators (see 
\S\,\ref{sec:pms}), it could strongly modify the main-sequence rotation profile
of slow rotators;
\item At the end of the sub-giant branch (\ie, at the boundary of the
Hayashi track), angular momentum redistribution is once more quite efficient and
could totally change the angular momentum profile that has been established during
the main sequence and the sub-giant phase contraction;
\item During the early-AGB, IGW generation again becomes substantial: mixing can be
expected to occur both at the convective boundary, due to the presence of an SLO and at the
hydrogen-burning-shell boundary, due to the enhanced damping of waves caused by the 
increase in the Brunt-V\"ais\"al\"a frequency.
\end{itemize}
Stellar models including meridional circulation, shear turbulence, and wave-induced
mixing are underway to quantify this impact (Charbonnel \& Talon, in preparation).

\begin{acknowledgements}
We thank the referee for her/his comments that helped us improve the 
manuscript. CC is supported by the Swiss National Science Foundation (FNS).
We acknowledge the financial support of the Programme National de
Physique Stellaire (PNPS) of CNRS/INSU, France.
Computational resources were provided by the R\'eseau qu\'eb\'ecois de 
calcul de haute performance (RQCHP).
\end{acknowledgements}

\end{document}